\documentclass[namedreferences,hyperref,optionalrh]{spr-sola}
\usepackage{graphicx}  
\usepackage{amssymb}      
\usepackage{color}
\newcommand{\rev}[1]{{\textbf{#1}}}  

\usepackage{amsmath}

\chardef\us=`\_

\begin{document}

\begin{frontmatter}
\title{Linear Polarimetry of the Solar Corona during the Total Solar Eclipse of 8 April 2024: Multi-Band Observations and HDR Imaging}

\author[addressref=aff1,email={suprit@iitd.ac.in}]{\inits{S.}\fnm{Suprit}~\snm{Singh}\orcid{0000-0002-7926-3174}}
\author[addressref=aff2,email={paras.sharma@stonybrook.edu}]{\inits{P.}\fnm{Paras}~\snm{Sharma}\orcid{0000-0002-6909-205X}}
\author[corref,addressref=aff3,email={bhartiarora612@gmail.com}]{\inits{B.}\fnm{Bharti}~\snm{Arora}\orcid{0000-0002-1360-4853}}
\author[addressref={aff1},email={Shivam.Burman@physics.iitd.ac.in}]{\inits{S.}\fnm{Shivam}~\snm{Burman}\orcid{0000-0003-0907-4308}}
\author[addressref={aff4},email={amchaitanya1708@gmail.com}]{\inits{C.}\fnm{Chaitanya}~\snm{Gandhi}}

\address[id=aff1]{Indian Institute of Technology Delhi, Hauz Khas, New Delhi, Delhi 110016, India}
\address[id=aff2]{Department of Physics and Astronomy, Stony Brook University, Stony Brook, NY 11794-3800, USA}
\address[id=aff3]{Jaypee Institute of Information Technology, A 10 Sector 62, Noida 201309, India}
\address[id=aff4]{St. Stephen's College, University Enclave, North Campus, Delhi 110007, India}

\runningauthor{S Singh et al.}

\begin{abstract}
High-Dynamic-Range (HDR) multi-band linear polarimetry of the solar corona during total solar eclipses provides a vital ground-based diagnostic tool for mapping plasma morphology and magnetic topology in the inner and middle corona. Extracting reliable polarimetric signals requires rigorous observational calibration, spatial co-alignment, and noise reduction via stacking. We report optical linear polarimetric observations acquired during the 8 April 2024 total solar eclipse from Dallas, Texas, USA, near solar maximum. Using a full-frame color CMOS detector paired with an 80~mm refractor, multi-band Red, Green, and Blue (RGB) images were recorded simultaneously across three polarizer orientations ($-60^\circ$, $0^\circ$, and $+60^\circ$) using a 9-step exposure bracket ($1/800\text{~s}$ to $1/3\text{~s}$). We present an end-to-end data-reduction pipeline incorporating: 
(i)~a novel astrometric registration framework that reconstructs Sun-centric solar-disk coordinates via lunar-limb fitting, catalog star plate-solving, and ephemeris displacement calculations, achieving sub-pixel alignment; 
(ii)~a photometrically linear exposure-normalization HDR image synthesis; and 
(iii)~Multi-scale Gaussian Normalization (MGN) coupled with Polarization-to-RGB chromatic mapping. 
The resulting structurally enhanced polarization composites unravel a complex solar-maximum coronal topology characterized by helmet streamers and fine-scale striations traceable out to $\sim3\,R_\odot$. This framework establishes a high-resolution, photometrically calibrated dataset ready for subsequent polarized brightness ($pB$) and electron density inversions.
\end{abstract}


\end{frontmatter}

\section{Introduction}
     \label{S-Introduction} 

The elusive solar corona has three primary components, the K-corona and the F-corona arising due to the scattering and the E-corona which consists of emission lines from the highly ionized solar atmosphere \citep{Golub_Pasachoff_2009}. The K-corona, which typically dominates within $\sim$1--3 solar radii ($R_\odot$), is produced by Thomson scattering of photospheric radiation by free electrons in the coronal plasma. Owing to the anisotropic nature of the scattering geometry, the K-corona is linearly polarised, with the polarization direction oriented tangentially to the solar limb (see \citealt{Billings_1966, Inhester_2015, Schad_2022}). Since this polarization state depends on spatial locations of the scattering electrons, the local electron density gradients and the underlying magnetic topology, polarimetric observations of the solar corona provide an important diagnostic tool for investigating the coronal structure and dynamics \citep{Howard_2009, Habbal_2011, Hanaoka_2012SoPh..279...75H, Bemporad_2020, Hanaoka_2021SoPh..296..158H, Edwards_2023, DelZanna_2023}. In contrast, the nearly unpolarised F-corona arises from sunlight scattered by the interplanetary dust particles; it dominates the signal beyond $\sim$ 2.5--3 $R_\odot$ and gradually fades into the zodiacal light at larger solar-centric distances (\citealt{Koutchmy_1985, Boe_2021}).

Observations of the inner/middle K-corona are extremely challenging under normal conditions because of overwhelming brightness of the solar photosphere, atmospheric sky scattering and significant stray light scattering within the observing instruments. Even in space-based telescopes, internal instrumental scattering poses a major challenge for inner coronal observations, as seen with the COR1 coronagraph onboard the STEREO spacecraft \citep{howard2008sun}. This has been mitigated to a great extent in a unique, recently launched Proba-3 concept \citep{Lamy_2010, Zhukov_2025}, wherein two spacecraft flying in formation allows achieving a better occultation and a significant reduction of the instrumental scattered stray light in the observations of the inner corona. Total Solar Eclipses (TSEs) offer a unique astrophysical laboratory to observe the solar corona right from near the solar disk $\sim 1R_\odot$ to out into the space $\sim 10R_\odot$. 

Imaging the solar corona presents significant observational challenges due to its extreme dynamic range. The combined brightness of the K, F, and E corona plummets from a normalized value of $1$ at the solar limb down to $10^{-3}$ at $2\,R_\odot$, and drops below $10^{-6}$ beyond $5\,R_\odot$. Standard consumer CMOS sensors cannot capture this steep gradient in a single frame; short exposures preserve the bright inner corona but lose faint outer structures, while long exposures saturate the inner regions. Overcoming this hurdle requires High Dynamic Range (HDR) imaging through exposure bracketing -- combining a sequence of short to long exposures into a single composite. Because total eclipses are brief, this strategy necessitates a camera with a high frame rate to cycle through brackets rapidly and repeatedly. Capturing multiple identical exposure sets enables frame stacking to boost the signal-to-noise ratio (SNR). Stacking is essential for faint outer regions, where low photon counts are otherwise dominated by photon shot noise and high thermal dark current from uncooled sensors operating in warm daytime conditions.

As capturing the vast intensity gradient necessitates exposure bracketed sequences, an accurate image co-alignment routine is a critical step before stacking in the eclipse-based HDR imaging of the corona and its scientific analysis. To achieve this, most previous studies have relied on phase-correlation methods, direct coronal feature matching, or stellar alignment techniques for image registration \citep{Druckmuller_2009, Bemporad_2020, Hanaoka_2021SoPh..296..158H, Edwards_2023}. However, feature-tracking methods can be compromised by the intrinsic evolution of the corona during the eclipse window, while stellar alignment depends heavily on the availability and detection of background stars against the inner-coronal glare. Stacking the eclipse images in a highly stable, solar-centric reference frame is therefore essential to avoid blending fine structures and to properly account for any spatio-temporal variations used for inferring coronal dynamics. Crucially, establishing this reference frame requires precisely locating the center of the solar disk, a formidable challenge during totality when the Sun is completely obscured by the Moon. 

Once a stable, co-aligned dataset is achieved, advanced preprocessing and filtering techniques are required to reveal the underlying fine-scale structures obscured by the steep radial intensity gradient. A variety of image-processing pipelines have been developed over the years to compress this dynamic range and enhance contrast (see, for example, \citealt{Byrne_2012, Druckmuller_2013, Morgan_2014, QIANG_2020, Patel_2022, Seaton_2023, DBSeaton2024SoPh..299...79S, Citizen_cate_2024}). Among the widely used approaches is the Multiscale Gaussian Normalization (MGN) technique \citep{Morgan_2014}, which enhances coronal fine structures by locally normalizing image intensities over multiple Gaussian spatial scales. This method has proven particularly effective in revealing filamentary structures, streamers, and faint coronal features while preserving overall morphology. In addition to enhancing the structural details, recent studies have also explored novel approaches for visualizing the polarised nature of the K-corona. Building upon these methodologies, \citet{Deforest_2022ApJ...927...98D, RPatel2023} introduced a visualization technique in which linear polarization observations obtained at three polarization orientations ($-60^\circ$, $0^\circ$, $+60^\circ$) are mapped directly onto the RGB colour channels. 

In this work, we present a comprehensive framework addressing each of these instrumental and analytical hurdles -- spanning acquisition, co-alignment, preprocessing, and visualization -- using multi-band linear polarimetry captured during the April 8, 2024 total solar eclipse. First, regarding acquisition, we leverage state-of-the-art commercially available mirrorless\footnote{The latest offerings from camera manufacturers are the mirrorless cameras which are essentially DSLRs sans the mirror and have an electronic viewfinder. The low latency displays on the back of the camera provide a Live View that is extremely helpful in quickly checking focus during the exposures.} cameras selecting models that optimally balance our requirements of high dynamic range, high frame rates, and the spatial resolution dictated by the aperture of our telescope. Also these are color cameras having a Bayer filter array on top of the pixels of the sensor allowing a simultaneous recording of the intensity data into Red, Green, and Blue (RGB) color channels. These can be separated and hence allow us to obtain the polarized brightness of the solar corona in the three channels. The multi-band data is useful in separating the K-corona, F-corona, and sky, which have different spectral characteristics \citep{Hanaoka_2021SoPh..296..158H}. Second, for image registration, we adopt a novel astrometric approach in which individual eclipse frames are co-aligned using dynamically reconstructed solar-centric coordinates derived from lunar-limb fitting, celestial coordinate transformations, and Sun-Moon positional calculations. This method places all frames in a common Sun-centred frame and enables consistent co-addition of multi-exposure polarimetric observations for subsequent HDR reconstruction. Finally, we implement advanced structural filtering and polarization-to-color mapping techniques to visualize the polarization signature of the solar corona at the solar maximum.

The remainder of this paper is organized as follows: Section \ref{sec:obs} details our multi-band instrumentation setup and eclipse observations. Section \ref{sec:analysis_and_results} presents our data analysis pipelines and results. 
A comprehensive scientific evaluation of these findings is provided in Section \ref{sec:discussion}, before concluding in Section \ref{sec:conclusions}.

\section{Observations} \label{sec:obs}

\begin{figure*}[htbp]
    \centering
    \includegraphics[width=0.99\textwidth]{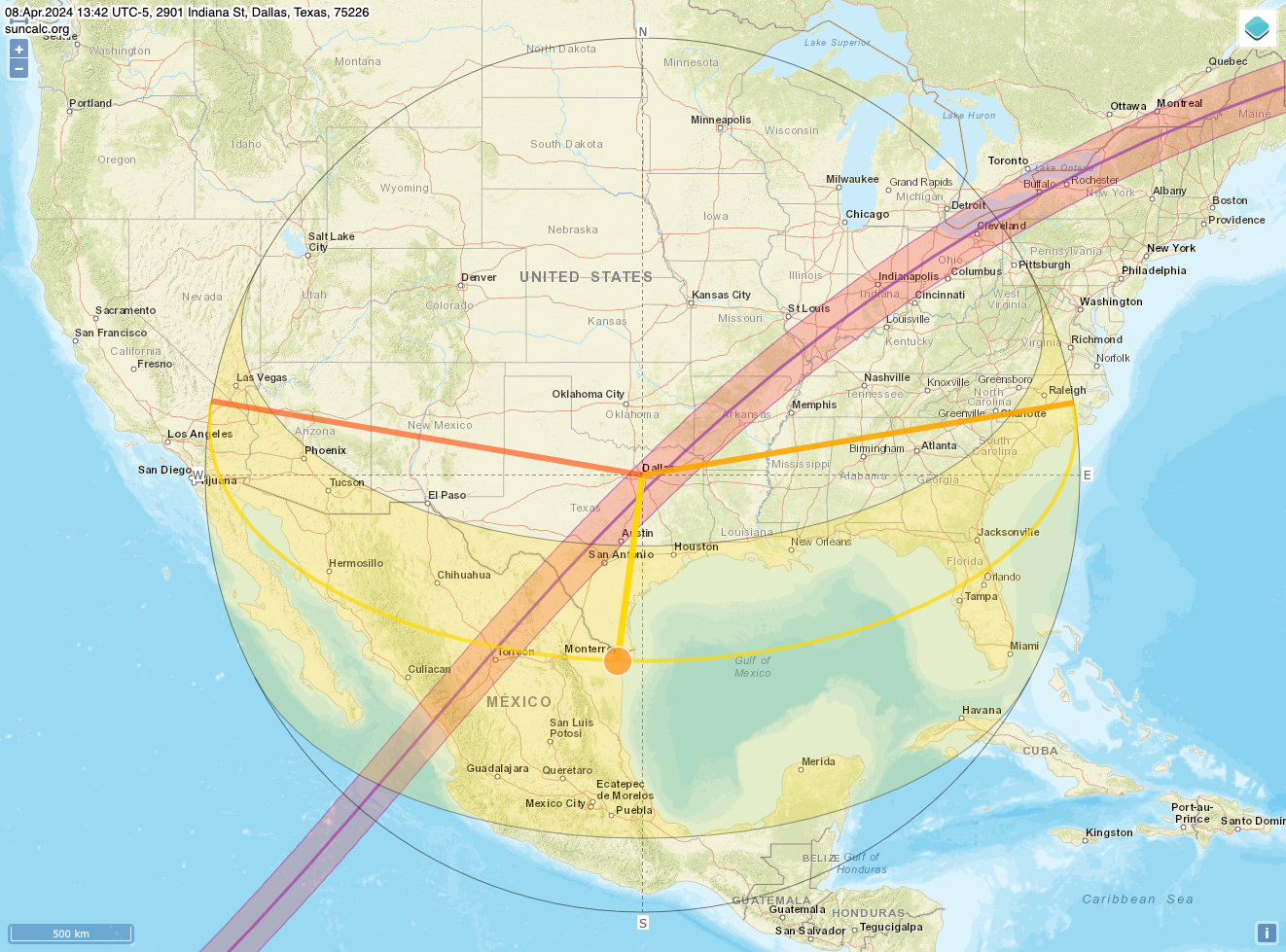}
    \caption{The figure shows the path of totality the April 8, 2024 total solar eclipse over the contiguous United States. The distinct line running directly through the middle of the purple/red band marks the centerline of the eclipse. Observers located exactly on this line experienced the maximum possible duration of darkness (totality). Stepping away from the central line reduces the the duration of the eclipse. The width of the band represents the physical diameter of the Moon's darkest shadow (the umbra) as it projects diagonally onto Earth's curved surface. Observers outside the band experience a partial solar eclipse. Our azimuth circle is shown centered on the data collection site in Dallas, Texas during totality (13:42:00 CDT). The dark yellow line towards the East points gives the azimuth of Sunrise, and the orange line towards the west gives the azimuth of Sunset, and the position of the Sun (orange circle) corresponds to its azimuth (slightly past the meridian) at the maximum totality, and the yellow arc denotes the Sun's trajectory across the sky on the day of the eclipse. Map generated via \href{https://www.suncalc.org/\#/32.7858,-96.7821,5/2024.04.08/13:42/1/3}{suncalc.org}.}
    \label{fig:apr_8_2024_Eclipse_path}
\end{figure*}

Our observations of the solar corona were conducted during the total solar eclipse of Monday, April 8, 2024, from Dallas, Texas, USA (\href{https://maps.app.goo.gl/mVUN3vocjgtSvWf5A}{$32^\circ47'08.34''$\,N, $96^\circ46'57.36''$\,W}). The totality spanned \href{https://www.timeanddate.com/eclipse/in/usa/dallas?iso=20240408}{3 minutes and 52 seconds} (13:40:43 -- 13:44:35 CDT) during which several sequences of images were acquired. Figure~\ref{fig:apr_8_2024_Eclipse_path} shows the path of the total solar eclipse over the contiguous United States centered on our data collection site.

\begin{figure}[htbp]
    \centering
    \includegraphics[width=0.99\textwidth]{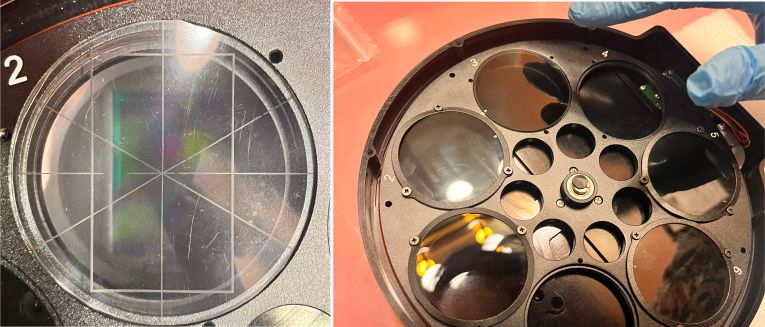}
    \caption{\textit{Left panel:} A printed transparency that aided the aligning of the polarization filters in the filter wheel as per the adopted reference frame ($-60^{\circ}$, $0^{\circ}$, and $+60^{\circ}$) with respect to the horizontal axis of the camera sensor visible underneath. \textit{Right panel:} Mounting of the polarization filters in the ZWO 2-inch filter wheel.}
    \label{fig:filters}
\end{figure}

\begin{figure}[htbp]
    \centering
    \includegraphics[width=\textwidth]{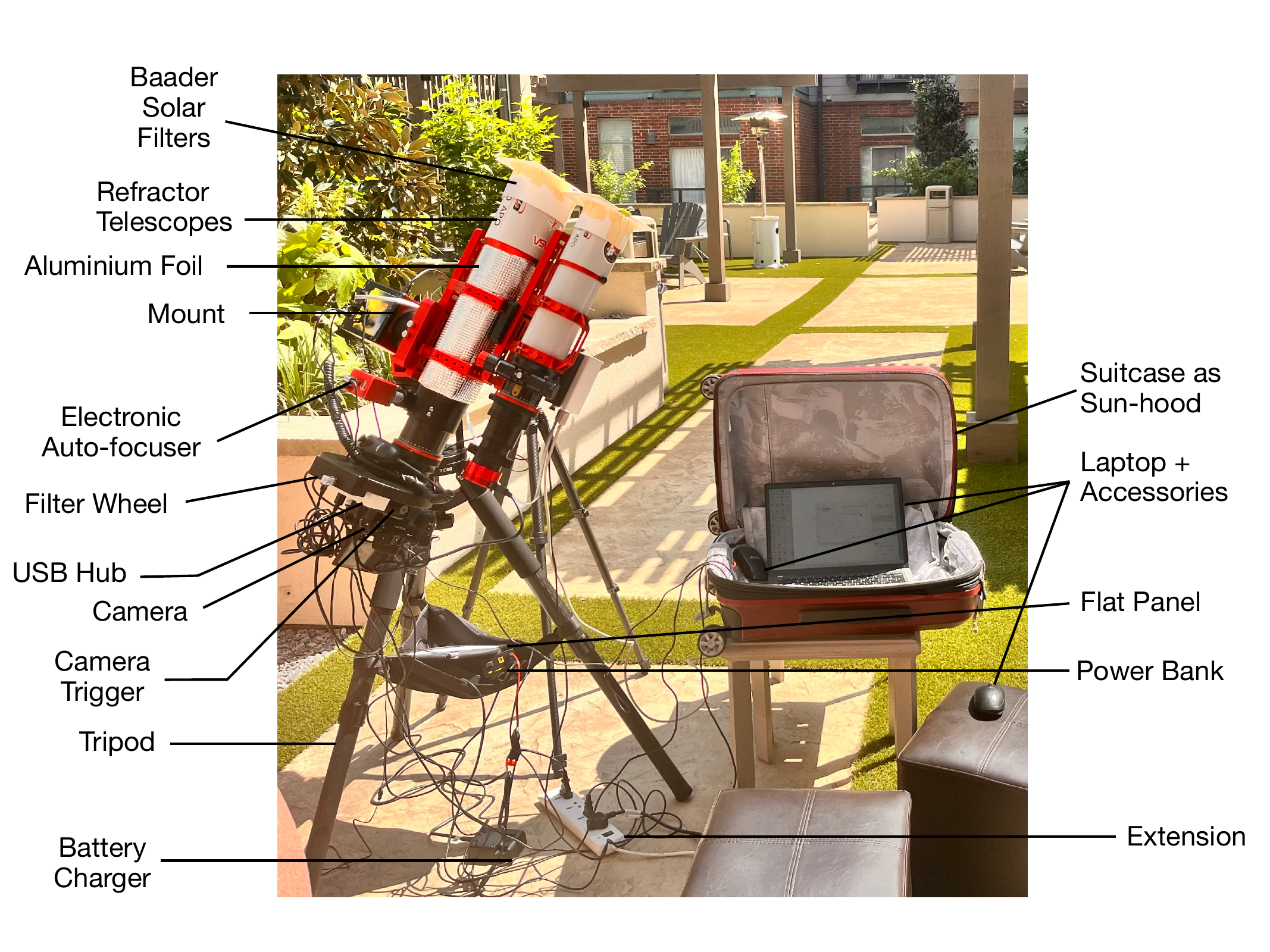}
    \caption{The observational setup employed for the polarimetric observations of the solar corona during the total solar eclipse.}
    \label{fig:equipment}
\end{figure}

The observations were performed using a ZWO FF80 80\,mm aperture f/7.5 refractor telescope, paired with a Nikon Z7II mirrorless camera featuring a 45.7\,MP BSI CMOS full-frame sensor with RGGB Bayer array to record the three RGB channels simultaneously. Baader OD5 filters were utilized during the partial phases to obtain full-disk images for photometric calibration. The telescope tracked the Sun on an equatorially configured ZWO AM5 mount and guided continuously using a Hutech Hinode Solar Guider. The optical train also incorporated a 2-inch ZWO filter wheel that was fitted  with three linear polarizing filters (each 50\,mm in diameter) aligned at $-60^\circ$, $0^\circ$, and $+60^\circ$ with respect to the horizontal axis of the camera sensor (Figure~\ref{fig:filters}). This physical alignment was carried out by printing the adopted reference frame on a transparent sheet as shown in the left panel of Figure~\ref{fig:filters}, which was subsequently used to mark the reference axes. Figure~\ref{fig:equipment} and Table~\ref{tab:equipment} summarize the components of our observational setup. With this optical train, the nominal plate scale was $1.49''/\mathrm{pixel}$ with the field of view of $3.417^{\circ} \times 2.278^{\circ}$.

\begin{table}{!ht}
\caption{List of equipment used for polarimetric observations during the 2024 TSE.}
\label{tab:equipment}
\begin{tabular}{ll}
\hline
Item & Manufacturer \& Model Number \\
\hline
Camera & Nikon Z7II, $35.9 \times 23.9$ mm, $8256 \times 5504$ pixels,\\
& $4.35 \mu$m pixel pitch full-frame RGGB sensor\\
Telescope & ZWO FF80 80~mm \textit{f}/7.5 APO refractor\\
Mount & ZWO AM5 Harmonic Go-To mount in Equatorial Configuration \\
~ & ZWO TC40 carbon fiber tripod and pier extension PE160\\
Laptop & Lenovo Thinkpad X13 Gen3, i5-1240P, 16GB RAM, 1TB SSD\\
Solar Filter & Baader Planetarium AstroSolar OD5 \\
Guider & Hutech Hinode Solar Guider\\
Diffuser & Aurora Flatfield Panel D=160 mm, with 12 V Inverter \\
Filter Wheel & ZWO $7\times 2$ inch EFW  \\
Focuser & ZWO EAF + Handcontroller\\
Polarizers  & Edmund Optics Linear Glass Polarizers ($\phi 50$mm)\\
GPS & VK-162 \\
Optical adapters & William Optics Z mount to M48 \\
Bubble Level & PS Align Pro App on iPhone 13 Mini \\
Compass & Suunto MC-2G Global Compass \\
Camera Trigger & LRTimelapse Pro Timer 3 \\
Power Supplies & Talentcell 12V 8300mAh YB1208300-USB\\
Storage  & Lexar CFexpress Type B USB 3.2 Gen 2x2 Reader,\\ 
      & Samsung T7 Shield 1TB SSD,\\ 
      & ProGrade Digital 128GB CFexpress 2.0 Memory Card\\
Miscellaneous & 3M Velcros, Cables, QZ Active Powered USB Hub\\
\hline
\end{tabular}
\end{table}

\begin{figure*}[htbp]
    \centering
    \includegraphics[width=0.99\textwidth]{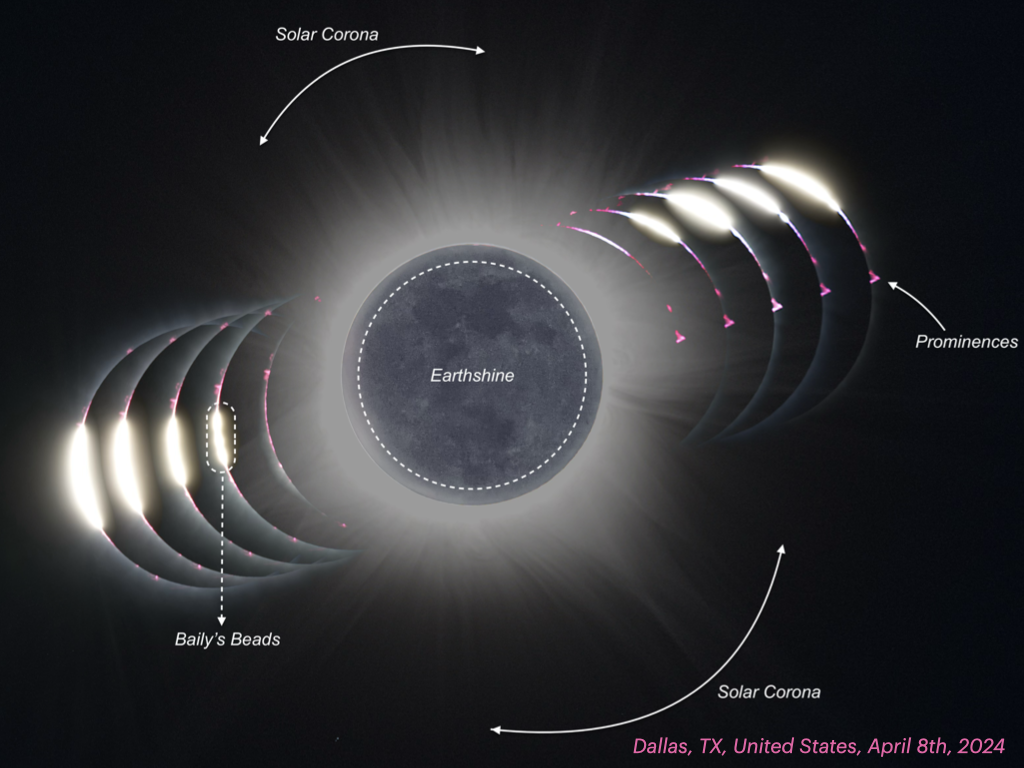}
    \caption{A multi-exposure time-composite of the April 8, 2024, total solar eclipse captured from Dallas, Texas, USA. The composite showcases the transition from the diamond ring and Baily's beads to the fully revealed solar maximum corona. Prominent magenta solar prominences are visible along the limb, while deep-exposure stacking resolves faint Earthshine structures on the lunar surface. These images were taken without polarisers.}
    \label{fig:progression_image}
\end{figure*}

The white-light solar corona exhibits an extremely large dynamic range in brightness, extending from the bright inner corona near the solar limb to the faint middle coronal structures several solar radii away \citep{West_2023}. As discussed in the introduction, capturing this broad intensity range within a single exposure of our detector, at the signal-to-noise ratio required here, is not feasible, as long exposures lead to saturation of the bright inner corona, whereas short exposures suffer from low SNR due to the weak signal from the outer regions. Consequently, a sequence of observations over multiple exposure times is necessary to simultaneously preserve both bright and faint coronal features. To visually contextualize the morphological state of the solar corona and illustrate this extreme dynamic gradient, a separate sequence of unpolarized broadband images was captured during the partial and total phases. Figure~\ref{fig:progression_image} presents a time-composite reconstructed from this data, tracking the progression of Baily's beads, prominent solar prominences along the limb, and faint Earthshine features on the lunar disk.

For our primary polarimetric experiment, observations were taken at three linear polarization orientations corresponding to \(-60^{\circ }\), \(0^{\circ }\), and \(+60^{\circ }\). For each polarizer position, image sequences were recorded over a wide range of exposure times. The exposure time series consisted of \(1/800\), \(1/400\), \(1/200\), \(1/100\), \(1/50\), \(1/25\), \(1/13\), \(1/6\), and \(1/3\) seconds. The total shutter time for the sequence was $\approx 0.63$\,s per bracket (from the actual exposure durations, Section~\ref{subsec: hdr}). Multiple frames (typically around five) were acquired at each exposure setting to improve the SNR by stacking the images subsequently. A bracket of nine exposures took about 2\,s and was repeated every 2--3\,s; the frames of the three polarizer orientations were taken between 13:40:58--13:41:32, 13:41:55--13:42:20 and 13:43:04--13:43:16 CDT. In addition to the observations taken during totality, a set of 43 full-disk solar images were acquired post eclipse for calibration and estimation of the absolute solar brightness. 


\textit{Frame selection:} The observed dataset consists of 136 coronal images spanning the three polarization orientations and multiple exposure levels. Every frame was inspected visually as well as quantitatively before stacking.  We took the criterion that the fitted lunar radius (Section~\ref{sec_align_images}) has to lie within 1-$\sigma$ of the mean radius over all frames. Thus, 6 of the 136 frames were rejected, and the remaining 130 frames were selected for stacking across the three polarizations at different exposure times.

Because our observations were recorded using a color sensor, data was captured across the RGB channels simultaneously. The raw frames were calibrated using standard image reduction procedures—including dark-frame subtraction, flat-field correction, and bilinear debayering. No normalization and white balance corrections were made. The individual calibrated R, G, and B channel images were subsequently separated and archived for independent spectral analysis. 

\section{Analysis and Results}
\label{sec:analysis_and_results}

\subsection{Synthetic Luminance Channels}
\label{subsec:synthetic_lum}

To extract the total intensity profile and structural details captured across the three spectral bands, a single-channel synthetic luminance frame ($L$) was constructed from the calibrated RGB channels using the standard, weighted luminosity relation \citep{iec1999srgb}:
\begin{equation}
L = 0.2126R + 0.7152G + 0.0722B 
\end{equation}
where $R$, $G$, and $B$ denote the calibrated red, green, and blue channel intensities, respectively. This relation collapses the multi-band data into a single high-resolution intensity map while preserving the underlying brightness variations and fine-scale structures of the corona. In this work, all subsequent structural processing, HDR reconstruction, and polarimetric analyses of the stacked images are performed exclusively using this single-channel luminance data, while independent polarization analyses of the individual spectral channels are reserved for a future study.

\subsection{Image Alignment}\label{sec_align_images}

Precise image co-alignment prior to stacking is essential to maximize the signal-to-noise ratio (SNR) across multi-exposure eclipse sequences. Existing registration techniques in the literature typically rely on phase correlation of coronal structures \citep{Druckmuller_2009}, spatial cross-correlation of fine coronal features \citep{Hanaoka_2021SoPh..296..158H}, or feature alignment using background reference stars \citep{Bemporad_2020, Edwards_2023}. However, each approach introduces physical constraints: feature-tracking assumes spatial stability of the dynamic corona across the exposure cadence, whereas star- or lunar-based tracking must account for the differential topocentric motion between the Sun, Moon, and background stars. To prevent spatial blurring and preserve physical polarimetric signatures, registration must be executed within a Sun-centric reference frame tied directly to the physical center of the solar disk. Because the solar disk is fully occulted during totality, determining these unobservable center coordinates is non-trivial. In this section, we present a Sun-centric alignment pipeline developed to address these requirements.

Our alignment approach relies on the precise knowledge of the direction of the Celestial North in the image frame (Solar North subsequently defines the polarization reference) to find the center of the hidden solar disk, and map spatial translations and rotations accurately across all images.

\begin{figure*}[htbp]
    \centering
    \includegraphics[width=0.99\textwidth]{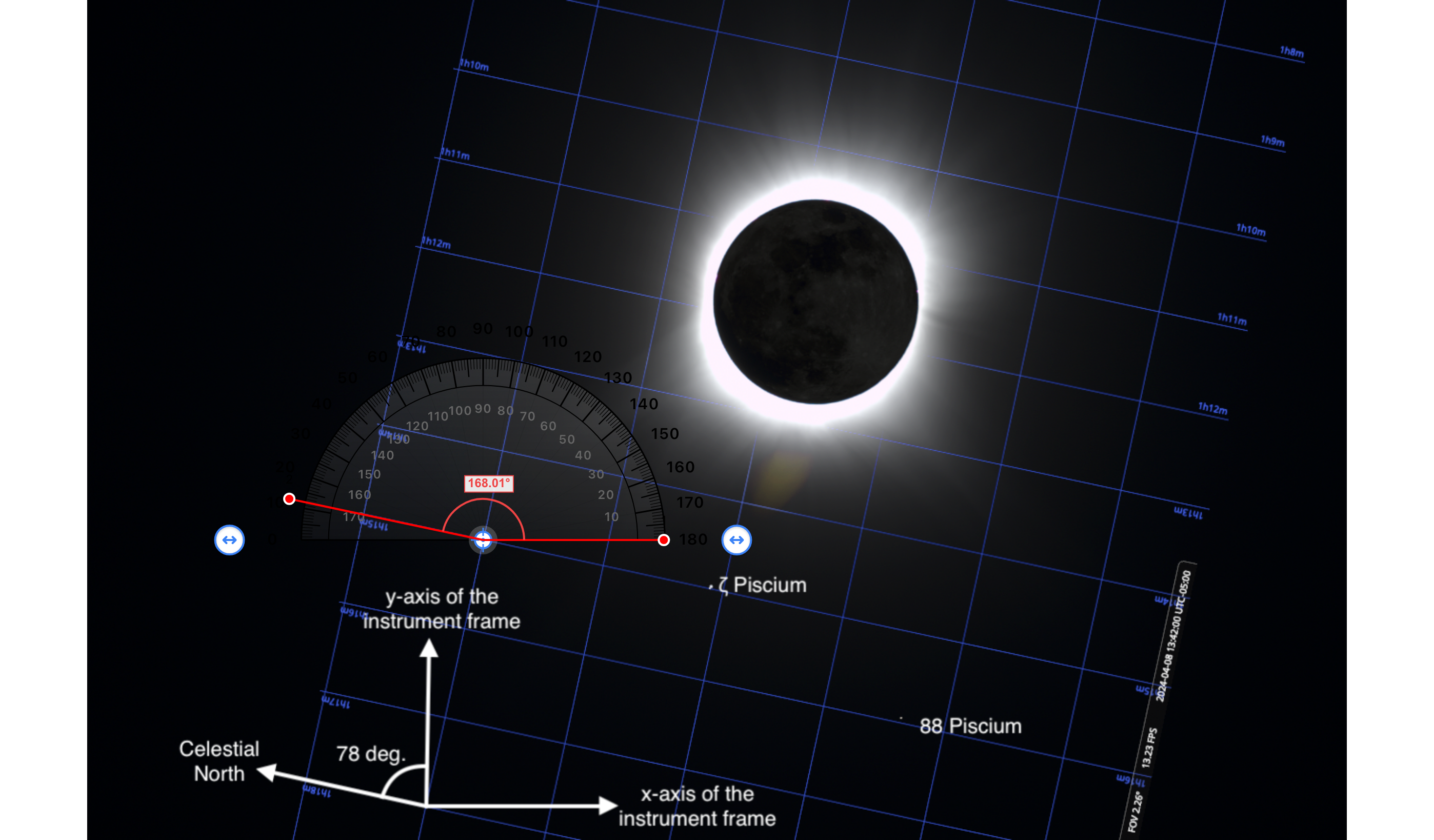}
    \caption{Figure shows Stellarium generated image (with equatorial grids) at the time of totality overlaid on top of one of our image. We found the celestial north by aligning the stars $\zeta$ Piscium (\rev{a double star}) and 88 Piscium in the field. The y-axis here is in instrument frame, that is, the frame of our captured images and thus the celestial north is at an angle of 168$^\circ$ from the instrument frame's x-axis. The protractor was overlaid with the \href{https://onlineprotractortool.org/}{https://onlineprotractortool.org/} to measure the angle.}
    \label{fig:finding_celestial_north}
\end{figure*}

To establish this reference frame, the orientation of Celestial North in the image coordinate system was determined using Stellarium \citep{Zotti_Hoffmann_Wolf_2021} by matching the simulated field of view with the exact geographic coordinates, date, and timestamps of our observations. The Stellarium rendering was overlaid onto a $1/3\,\text{s}$ exposure frame and manually aligned against the observed background star pair ($\zeta$~Piscium and 88~Piscium). Superimposing the equatorial coordinate grid revealed that the Celestial North vector was oriented at a fixed angle of $168^\circ$ counter-clockwise relative to the positive horizontal $x$-axis of our image frame (see Figure~\ref{fig:finding_celestial_north}). To align the data with the solar rotation axis, the solar position angle ($P$-angle) was computed using the SunPy Python package \citep{sunpy_community2020} (\texttt{sunpy.coordinates.sun.P}, evaluated at mid-totality, 18:42:00\,UTC). On the date of observation, the solar North direction was found to be offset by $-26.24^\circ$ relative to Celestial North.

With the knowledge of celestial and solar North directions in our frames, the precise pixel coordinates of the unobservable solar center were mapped for each individual exposure using a three-step geometric pipeline:

\begin{enumerate}
    \item \textbf{Lunar Limb Detection:} A Circular Hough Transform (CHT) algorithm \citep{Hough_1962}, implemented via \texttt{scikit-image}, was applied to detect the lunar limb, yielding initial pixel coordinates for the Moon's center and its apparent radius.
      \item \textbf{Plate Scale and Sun--Moon Displacement Vector Computation:} The image plate scale was calculated for each frame as the ratio of the measured lunar pixel radius to its physical topocentric angular radius, computed using the \texttt{Skyfield} Python package \citep{Skyfield_Rhodes} with the JPL DE421 ephemeris. \texttt{Skyfield} was simultaneously used to derive the topocentric angular displacement vector of the Sun relative to the Moon and its orientation relative to Celestial North.

    \item \textbf{Sun-Centric Coordinate Mapping:} By converting the \texttt{Skyfield} angular displacement vector into pixel offsets via the derived plate scale, and rotating it to align with the $168^\circ$ Celestial North baseline (with East $90^\circ$ counter-clockwise from North, as on the sky), the absolute pixel coordinates of the solar-disk center were mapped for every frame.
\end{enumerate}

\begin{figure*}[htbp]
    \centering
    \includegraphics[width=\textwidth,height=0.73\textheight,keepaspectratio]{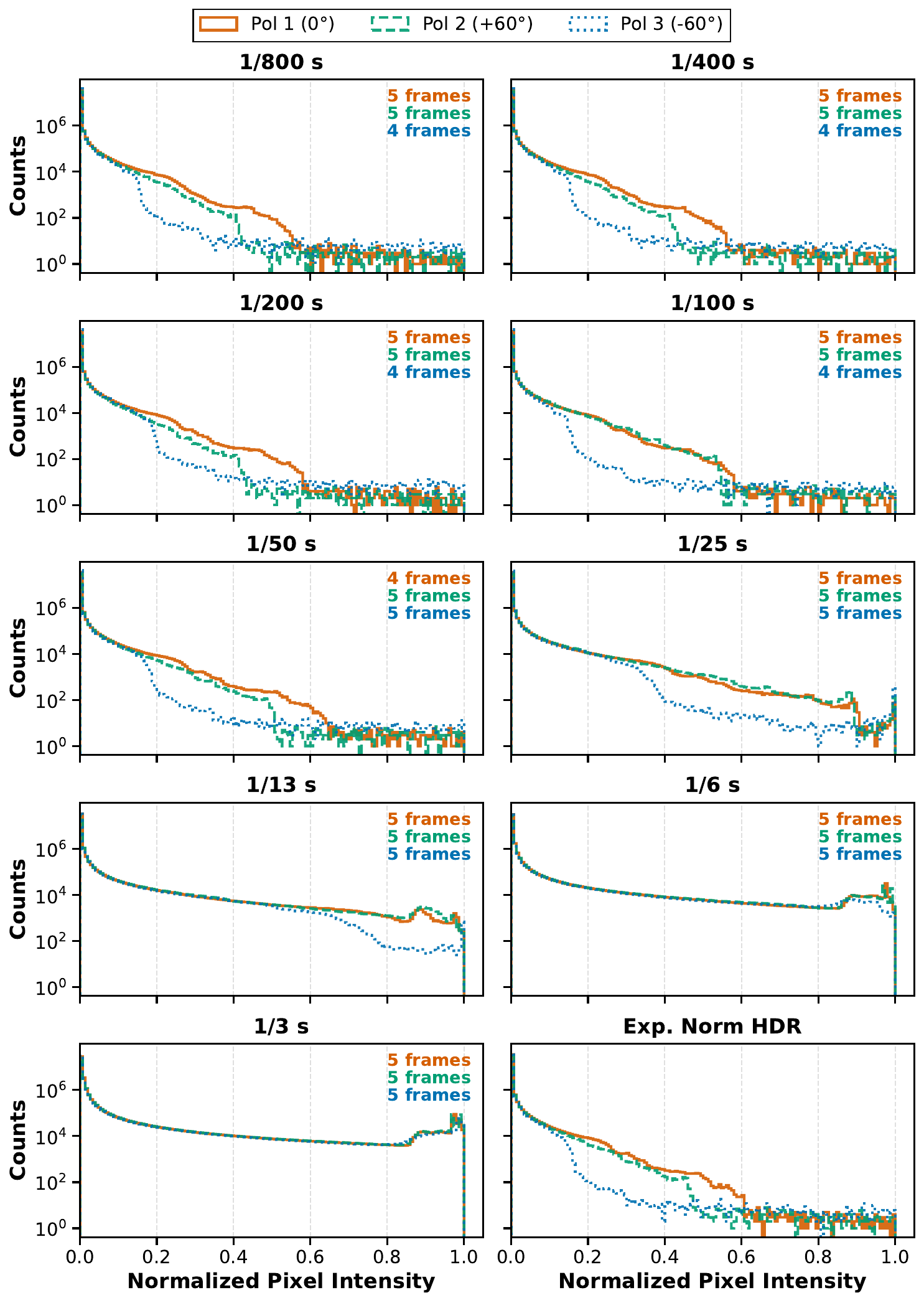}
    \caption{Normalised pixel intensity distributions across exposure times and the HDR reconstruction. The histograms display the number of pixels (logarithmic vertical axis) against the pixel intensity normalised to the maximum value within each respective channel (linear horizontal axis). Data are shown for the three linear polarization filter orientations: $0^{\circ}$ (vermillion, solid), $+60^{\circ}$ (bluish green, dashed), and $-60^{\circ}$ (blue, dotted). The first nine panels (left to right, top to bottom) show the distributions for individual stacked exposures, increasing from 1/800~s to 1/3~s. The coloured labels in these panels give the number of raw frames stacked to produce the image for each polarization state. The final panel shows the exposure-normalised HDR image.}
    \label{fig:histogram_comparison}
\end{figure*}

The longest exposures ($1/3\,\text{s}$) required one further alignment step because their bright inner corona is saturated and the lunar-limb fit is exposure dependent. Each $1/3\,\text{s}$ frame was registered to a $1/6\,\text{s}$ reference stack of the same polarizer orientation using the large-scale coronal brightness gradient. The displacement $\mathbf{d}$ was fitted by least squares using the first-order approximation:
\begin{equation}
\ln I_{1/3}(x, y) - \ln I_{1/6}(x, y) \simeq \ln c - \mathbf{d}\cdot\nabla \ln I_{1/6}(x, y)
\end{equation}
where $I_{1/3}$ and $I_{1/6}$ are the geometrically aligned frame and reference luminance images after Gaussian smoothing ($\sigma=4$\,pixels), $(x,y)$ are image coordinates in pixels, $\mathbf{d}=(d_x,d_y)$ is the relative pixel displacement, and $\nabla$ is the spatial gradient. The dimensionless factor $c$ accounts for the overall brightness ratio between exposure times. Following registration, each accepted frame was shifted by bilinear interpolation to a common Sun center at $(4128, 2752)$\,pixels, and frames within each polarizer and exposure sub-group were combined by average stacking.

Figure~\ref{fig:histogram_comparison} gives the pixel-intensity distributions of the resulting stacked images. The first nine panels illustrate how the sampled intensity range varies across the different exposure times, while the vermillion (solid), bluish-green (dashed), and blue (dotted) curves correspond to the three polariser orientations: $0^{\circ}$, $+60^{\circ}$, and  $-60^{\circ}$, respectively. The number of accepted frames contributing to each stacked image is indicated in each panel in Figure~\ref{fig:histogram_comparison}. Since the observations were obtained using three polarizer orientations across nine discrete exposure times, this procedure yielded a final set of 27 high-SNR stacked images ready for subsequent HDR reconstruction and polarimetric decomposition. 

Although a degree of saturation is evident in some of the longest exposures for certain polariser orientations, the ensemble of stacked images collectively provides a continuous coverage of the coronal intensity range. This highlights the importance of acquiring observations over multiple exposure times and provides the empirical foundation for the subsequent HDR reconstruction, enabling both bright and faint coronal structures to be preserved within a single composite representation. The histograms of the HDR images resemble those of the short exposures (1/800 to 1/200\,s), since both are normalised to the brightest inner corona; in the HDR images, however, the faint outer corona in the lowest bins comes from the 1/3\,s exposure rather than from noise-dominated short exposures.

\subsubsection{Error Estimates and Uncertainty Analysis}

Before moving further, we give the error estimates arising from all the steps of our pipeline below.

\begin{itemize}

    \item \textbf{Orientation and Ephemeris Uncertainties:} We quantify the uncertainty of the manual alignment step (upper bounded) by the lever-arm ratio of the point-spread-function (PSF) width to the inter-stellar baseline, $\Delta \theta \approx \text{FWHM} / D \approx 10\,\text{pixels} / 1500\,\text{pixels} \approx 0.0067\,\text{rad} \approx 0.4^\circ$. A least-squares fit of the catalogue positions of these stars -- using the flux-weighted position for the $\zeta$~Piscium binary -- to their measured image centroids confirmed this orientation and derived a plate scale of $1.5''$/pixel. Over our topocentric Sun--Moon separation ($26''\text{--}49''$, or $18\text{--}33$\,pixels), this $0.4\%$ scale difference shifts the inferred solar-disk center by at most $0.1$\,pixels. \\
      
    \item \textbf{CHT Error Budget:} Because the CHT operates on a discrete grid, integer parameter output for the center and the radius introduces a nominal discretization uncertainty of $\pm 0.5$\,pixels, which is independent of physical limb-fitting errors. The fitting errors in the radius and the center of the lunar disk (after removing the linear tracking drift) were $\approx 2$\,pixels and $\approx 2.6$\,pixels respectively. The error in the center of solar disk is dictated by the error in the center of the lunar disk directly, and by the error in the radius of the moon through the plate scale. The combined error in the center of the solar disc from both comes out to be $\approx 2.6$\,pixels dominated by the error in the center of the lunar disc.
  \end{itemize}

\subsection{HDR Image Construction by Exposure Normalisation Method}
\label{subsec: hdr}

We now turn to constructing an HDR image from the set of different exposures. The approach is a physically motivated exposure-normalised stacking method designed to preserve the relative photometric scaling of the coronal brightness, making it suitable for quantitative analyses such as polarimetry and coronal electron density estimation. We developed our own Python implementation of the technique using the libraries NumPy, SciPy, and Astropy, and following three core conventions:

\begin{enumerate}
    \item \textbf{Exposure times:} For reporting consistency, the exposures are identified throughout this paper by their nominal camera display labels (1/3, 1/6, 1/13, 1/25, 1/50, 1/100, 1/200, 1/400, and 1/800\,s). However, the commercial cameras show rounded values, and the actual physical shutter durations follow an exact geometric sequence $t_i = 2^{-n_i/3}$\,s for step indices $n_i \in \{5, 8, \ldots, 29\}$. This sequence yields precise integration times of 0.315, 0.157, 0.0787, 0.0394, 0.0197, 0.00984, 0.00492, 0.00246, and 0.00123\,s, ensuring that each consecutive exposure duration differs by an exact factor of two. We use these exact physical durations $t_i$ in the subsequent analysis. To verify this relative exposure scale in our coronal observations, we fitted the signal ratios of adjacent, unsaturated image stacks, obtaining empirical values between 1.99 and 2.02 across the 1/3\,s to 1/50\,s range, confirming linear relative shutter scaling. Additionally, laboratory calibration using a uniformly illuminated source and the camera gain set at ISO 100 mapped the sensor’s linear response regime and identified its upper saturation boundary (Appendix~\ref{app:linearity}, Figure~\ref{fig:sensor_linearity}).

    \item \textbf{Common normalisation:} Let $I_i^{(p)}(x,y)$ denote the stacked synthetic luminance at exposure duration $t_i$ for polariser orientation $p$. To evaluate pixel brightness on a standardized scale without altering raw count data, we define a single global normalization constant $I_{\max} = \max_{j,p,x,y} I_j^{(p)}(x,y)$, representing the peak intensity observed across all exposures, polariser angles, and spatial coordinates. Because $I_{\max}$ is used exclusively to evaluate mask thresholds, the raw calibrated values in the HDR numerator remain unscaled, preserving the exact intensity ratios across polariser orientations that carry the physical polarization signal.

    \item \textbf{Master luminance and selection mask:} For each exposure $t_i$, we construct a dimensionless master luminance map,
    
    \begin{equation}
        w_i(x,y) = \mathrm{max}_p I_i^{(p)}(x,y) / I_{\max},\nonumber 
    \end{equation} 
    by taking the spatial maximum across the three polariser channels relative to $I_{\max}$. The selection mask is required to isolate valid, linear detector measurements by excluding saturated pixels at high intensities and noise-dominated pixels at low intensities. Evaluating the mask $v_i(x,y)$ on $w_i(x,y)$ and applying it identically across all three polariser channels ensures that exposure transitions occur at the exact same spatial pixels across all channels, preventing boundary discontinuities and false polarization signals.
\end{enumerate}

To construct the HDR images of the solar corona from our 27 stacked frames, each image was normalised by its corresponding exposure time so as to preserve the intrinsic linear photometric scaling of the coronal brightness \citep{Hanaoka_2021SoPh..296..158H, Edwards_2023}. Let $I_i(x,y)$ denote the non-negative stacked synthetic luminance image (negative samples clipped to zero) for a given polariser orientation (with the index $p$ suppressed) at exposure time $t_i$, then the exposure-normalized image $I_i'(x,y)$ is defined as:
\begin{equation}
I_i'(x,y) = \frac{I_i(x,y)}{t_i},
\end{equation}
where $(x,y)$ represents the pixel spatial coordinates. For a linear detector, every $I_i'$ that is neither saturated nor dominated by noise is an estimate of the same count rate. The final HDR image $(I_{\mathrm{HDR}})$ for each polariser orientation was therefore obtained by combining only these well-exposed estimates, with the standard validity mask (modified at the longest and shortest exposures as described below),
\begin{equation}
I_{\mathrm{HDR}}(x,y) =
\frac{\sum_{i=1}^{N} v_i(x,y)\, t_iI_i'(x,y)}{\sum_{i=1}^{N} v_i(x,y)\, t_i},
\quad
v_i(x,y) =
\begin{cases}
1 & 2\% \le w_i(x,y) \le 80\%,\\
0 & \mathrm{otherwise},
\end{cases}
\end{equation}
where $N=9$ represents the total number of exposures for our case, ordered longest to shortest, and $w_i(x,y)$ is the dimensionless master luminance defined above; $v_i(x,y)$ is the binary validity mask. For the longest exposure, only the upper cut $w_i\le 80\%$ is applied, retaining the faint outer corona; for the shortest exposure, only the lower cut $w_i\ge 2\%$ is applied. The same mask and denominator apply to all polariser orientations. This is the exposure-time-weighted mean of the valid exposure normalized images \(I_i'\): at each pixel, an accepted image contributes with weight proportional to its exposure time \(t_i\). The adopted $2\%$ lower cut suppresses weak-signal read noise and calibration offsets, which the division by $t_i$ would strongly amplify for the shortest exposures; the adopted $80\%$ upper cut avoids near-saturation samples. Up to six exposures contribute in the inner corona, and beyond $\sim 2.2\,R_\odot$ the 1/3\,s exposure is predominantly used alone. This procedure results in three HDR images corresponding to each polarization position and also preserves the relative coronal intensity distribution over a broad dynamic range while simultaneously improving the SNR through multi-frame stacking.

\subsection{Multi-scale Gaussian Normalisation Processing}
\label{subsec:mgn}

To enhance the visibility and high-frequency contrast of fine-scale structures in the corona, the MGN technique developed by \citet{Morgan_2014} was applied to the final HDR intensity maps generated by the exposure normalization pipeline. The MGN algorithm systematically amplifies structural features across a broad range of spatial scales by locally normalizing image intensities using Gaussian kernels of varying widths. This multi-scale transformation effectively suppresses the massive bulk intensity gradient while simultaneously preserving large-scale morphological topology and bringing out faint, thread-like coronal structures.

To express the polarization angles with respect to the solar north, the Stokes parameters $I$, $Q$ and $U$ computed from the three HDR images (aligned at (\(-60^\circ\), \(0^\circ\), and \(+60^\circ\)) with respect to the horizontal axis of the camera sensor) were transformed to the intensities from the three polarizers at $0^\circ$, $+60^\circ$ and $-60^\circ$ from the solar north, which lies at $168^\circ - 26.24^\circ = 141.76^\circ$ from the image horizontal ($x$-axis). The implementation uses clockwise array angles ($x$ right, $y$ down), so the synthesis reference angle is $-141.76^\circ$. MGN (the \texttt{sunkit-image} implementation, \nolinkurl{sunkit_image.enhance.mgn}, with Gaussian widths of 10, 20, 40 and 80 pixels and the default parameters $k = 0.7$, $\gamma = 3.2$, $h = 0.7$ and equal weights) was applied to the three transformed HDR composites. False RGB colour channels were then assigned to these three polarization states, each channel carrying its fraction of the total intensity, and combined to form a single composite image, following visualisation approaches similar to those presented by \citet{RPatel2023} and \citet{Citizen_cate_2024}. The total intensity normalised to its maximum sets the brightness of the composite while the channel fractions set its colour. An additional display power of 0.4 is used for enhancing visual brightness. These display operations are separate from the linear HDR images to be used for quantitative measurements. This approach generates a single, intuitive composite image where spatial variations in color correspond to the shifts in the linear polarization vector. 
 
\begin{figure*}[htbp]
    \centering
    \includegraphics[width=1.0\textwidth]{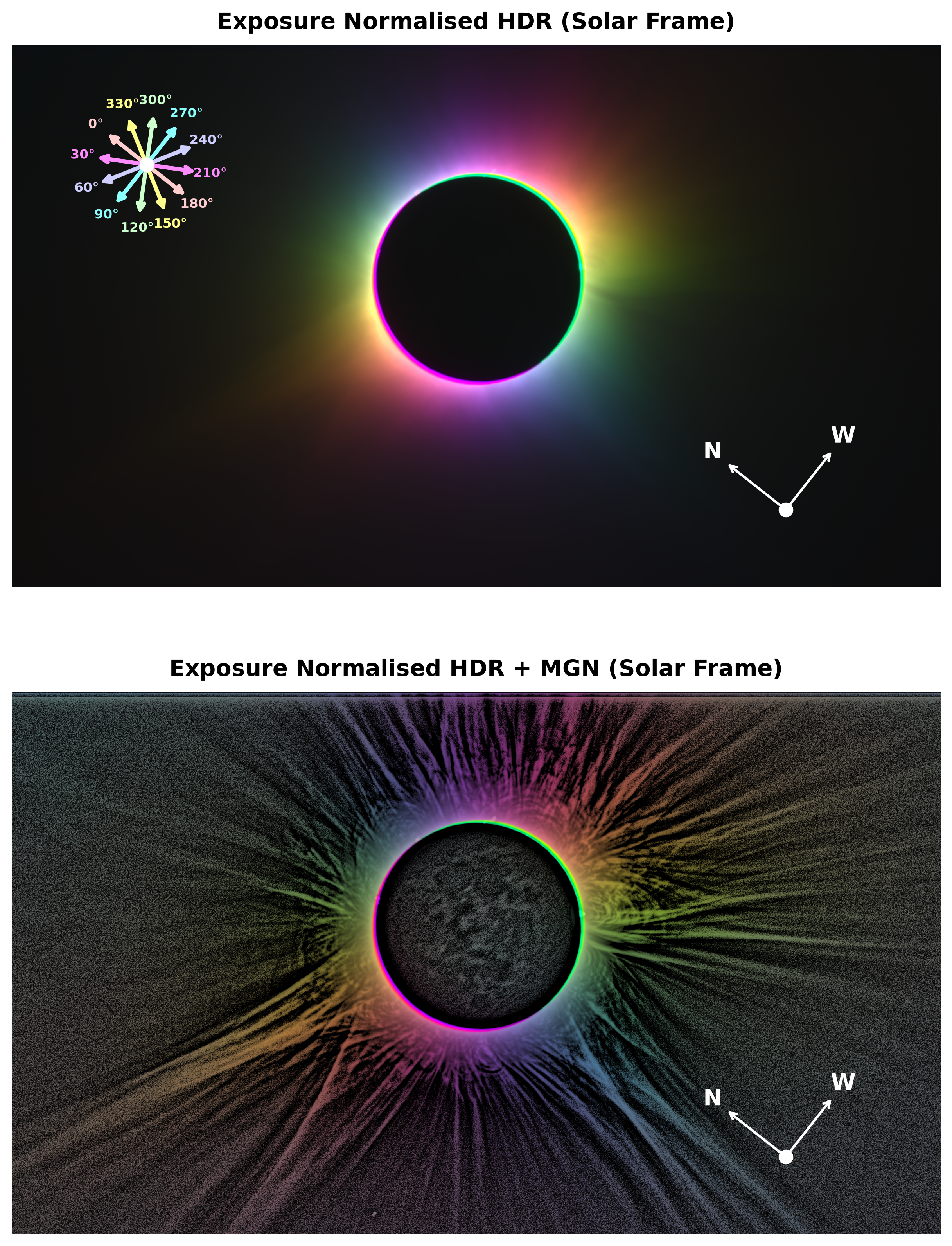}
    \caption{The upper panel shows the HDR image constructed using the exposure-normalisation technique, while the lower panel presents the same image after enhancement using the MGN technique. The intensities synthesized from the Stokes parameters at $0^\circ$, $+60^\circ$, and $-60^\circ$ with respect to the solar north are mapped to R, G, and B, respectively. The arrow marked \textit{N} denotes the direction of solar north relative to the image coordinate system. The colour wheel indicates the correspondence between colour and polarization state, where the polarization angle is measured from the solar north. The coloured rim along the lunar limb is not coronal. It is due to the motion of the Moon relative to the Sun between the first and the last polarizer orientation.}
    \label{fig:pol_1}
\end{figure*}

The upper panel of Figure~\ref{fig:pol_1} presents the colour HDR composite constructed from the exposure-normalisation method prior to MGN enhancement. The corresponding lower panel shows the same composite after MGN processing, bringing out the fine-scale coronal structures. 

We have presented an end-to-end pipeline with open-source components optimized for multi-band coronal polarimetry. This framework establishes the astrometric and data-reduction baseline for structural imaging, laying the groundwork for subsequent physical inversions.

We emphasize that the scope of this paper is limited to the observations, the Sun-centric alignment, and technique and construction of the HDR images; quantitative inversions of these multi-band datasets to extract the polarized brightness, coronal electron densities, and underlying magnetic fields are reserved for a forthcoming companion paper (Singh et al., in preparation).

\section{Scientific Evaluation and Discussion} \label{sec:discussion}

The high-dynamic-range linear polarization composites generated via the MGN pipeline show the morphological state of the solar maximum corona. As can be seen in the top panel of Figure~\ref{fig:pol_1}, the colour of the inner corona changes continuously with position angle and runs twice through the colour wheel around the Sun. This spatial color distribution directly maps the orientation of the linear polarization vector. The coloured rim along the lunar limb is not coronal: the Moon moved relative to the Sun by $\approx 28$ pixels between the first and the last polarizer orientation, so that the lunar limb lies at slightly different positions in the three Sun-centred channels. The MGN enhanced image also shows thread-like striations mapped between the dominant helmet complexes.

The structurally enhanced composite in the lower panel of Figure~\ref{fig:pol_1} reveals a complex, quasi-uniform distribution of helmet streamers extending outward across all heliographic latitudes. This distribution of large-scale structures is characteristic of the coronal environment near solar maximum, in contrast with the equatorially confined streamer belts observed during solar minimum. The combination of exposure normalized HDR and MGN processing retains fine structure from the lunar limb out to the middle corona ($r \sim 3\,R_\odot$). These images can provide a spatial baseline for subsequent inversions of polarized brightness and coronal electron density profiles.

\section{Conclusions}
\label{sec:conclusions}

We have presented multi-band linear polarimetric observations of the white-light solar maximum corona during the 8~April 2024 total solar eclipse from Dallas, Texas, captured across nine bracketed exposure levels with a commercial mirrorless color camera (Nikon Z7II) and an 80\,mm refracting telescope using a polarizer triplet aligned at $-60^\circ$, $0^\circ$ and $+60^\circ$ with respect to the horizontal axis of the camera sensor. The data acquisition lead to a simultaneous recording of polarization in the three RGB color channels.

The rest of the paper considers the crucial step of image co-alignment, registration, stacking and HDR image synthesis. We chose to work with synthetic luminance frames created from the calibrated light frame for this purpose while the separated RGB channels were archived for future analysis. 

To align the frames relative to the occulted solar disk, an astrometric Sun-centric registration pipeline was implemented integrating manual star plate-solving with Stellarium to determine the direction of the celestial north, Circular Hough Transform (CHT) lunar limb detection to obtain the lunar center and radius for determining the plate scale, and translating the ephemeris angular displacement vector between the Moon and the Sun using Skyfield with JPL DE421 to the image, determining the unobservable solar center with sub-pixel precision.

We confirmed the linearity in the response of our camera sensor, and incorporating masking thresholds on underexposed and saturated pixels, developed and implemented an exposure normalized linear HDR synthesis technique that preserved photometrically valid data across a wide dynamic range, bridging the inner corona ($1.1\,R_\odot$) and the middle coronal regions ($>3.0\,R_\odot$). To this, we then applied the Multi-Scale Gaussian Normalization (MGN) that enhanced the HDR composite revealing fine scale coronal features -- helmet streamers and striations.

A chromatic treatment to polarization was given by transforming the luminance channel polarization triplet intensities in the instrument frame to Sun centric frame aligned with the direction of the Solar North and assigning these transformed intensities to the RGB channels. This compositing rendered a false color image allowing a visualization of the coronal structure, brightness, and polarization in a single, high-contrast image.

The outputs generated here provide the required foundation for quantitative inversions of the polarized brightness, coronal electron densities, and magnetic field mapping which will be presented in a forthcoming study. The pipeline developed here also provides a fully reproducible framework that can be applied to future total solar eclipse observations.

\section*{Acknowledgements}
This research of S.S. was supported by the New Faculty Seed Grant and Professional Development Allowance (PDA) from the Indian Institute of Technology Delhi, India, and the Core Research Grant CRG/2021/003053 from the Science and Engineering Research Board, India. Research of S.B. is supported by the Institute Fellowship from the Indian Institute of Technology Delhi, India. We are also grateful to the anonymous reviewer for their constructive comments that greatly helped in improving the manuscript.

\section*{Data/Code availability}
The code and data used in this paper is available on reasonable request from the corresponding author.

\appendix

\section{Linearity of the CMOS Sensor in the Nikon Z7II Camera Used for Polarimetric Observations}
\label{app:linearity}

Precision astronomical polarimetry relies on a strictly linear relationship between incoming photon flux and registered analog-to-digital units (ADU). Non-linearities in sensor response or imprecise exposure timing introduce systematic errors into measured Stokes parameters. To characterize the response of the full-frame CMOS sensor in the Nikon Z7II camera body, a series of flat-field exposures was acquired across a broad range of integration times and evaluated against ideal linear response models.

\subsection{Exposure Time Calibration ($t_i = 2^{k/3}$)}

Standard digital camera shutter microcontrollers step exposure times in discrete $1/3$-stop intervals based on power-of-two timing rather than the nominal values displayed on the camera user interface. Using nominal display values (e.g., $1/3\text{~s} \approx 0.3333\text{~s}$) introduces systematic exposure timing errors up to $5.7\%$ relative to actual electronic timing ($2^{-5/3}\text{~s} \approx 0.3150\text{~s}$).

To maintain photometric accuracy, exposure durations $t_i$ were converted to exact power-of-two steps relative to a base $1.0\text{~s}$ exposure:
\begin{equation}
t_i = 2^{k/3} \quad \text{for } k \in \mathbb{Z}
\end{equation}

For example, sub-second sequences covering nominal exposures from $1/1000\text{~s}$ to $1/1.3\text{~s}$ correspond to $k \in \{-30, -27, -25, \dots, -1\}$, while longer exposures extend to $k = +8$ ($2^{8/3}\text{~s} \approx 6.350\text{~s}$).

\subsection{Bias Subtraction and Data Reduction}

Raw 14-bit FITS frames were reduced by isolating median frame statistics within the central region of interest. A physical master bias offset ($B=1008$ ADU) was derived independently from zero-second dark frames using the open-source astronomical software \textsc{Siril} \citep{Richard2024Siril} was subtracted from the raw median intensity $S_{\text{raw}}(t_i)$ to yield the net signal intensity $S_{\text{net}}(t_i)$:
\begin{equation}
S_{\text{net}}(t_i) = S_{\text{raw}}(t_i) - B
\end{equation}

The dataset was partitioned into two response regimes:
\begin{enumerate}
    \item \textbf{Unsaturated Linear Domain:} $t_i \le 1.0\text{~s}$ ($S_{\text{raw}} \le 12{,}048\text{~ADU}$)
    \item \textbf{Non-Linear \& Saturation Domain:} $t_i > 1.0\text{~s}$ ($S_{\text{raw}} > 12{,}048\text{~ADU}$)
\end{enumerate}

\subsection{Linear Response Modeling and Residual Analysis}

\begin{figure}[htbp]
    \centering
    \includegraphics[width=0.9\linewidth]{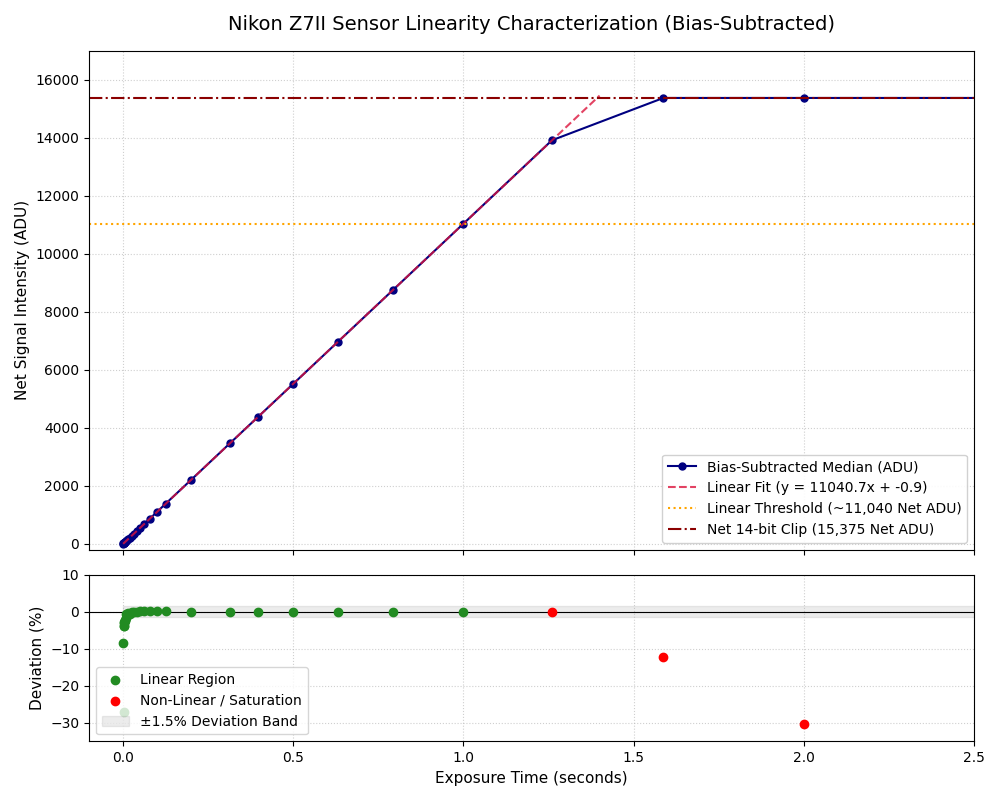}
    \caption{Laboratory linearity measurement of the Nikon Z7II at ISO 100. \textit{Top:} Bias-subtracted median signal versus exposure time, shown with the linear regression fit. \textit{Bottom:} Percentage deviation from the linear fit, highlighting non-linearity at low light levels due to noise and at high signal levels due to detector saturation.}
    \label{fig:sensor_linearity}
\end{figure}

\begin{table}[htbp]
\caption{Key linearity and sensor operational parameters for the Nikon Z7II CMOS sensor}
\label{tab:sensor_linearity}
\begin{tabular}{lll}
\hline
\textbf{Parameter} & \textbf{Symbol} & \textbf{Value} \\ \hline
Applied Bias Subtraction & $B$ & $1008$ ADU \\
Linear Scaling Factor & $m$ & Fit slope (ADU/s) \\
Residual Intercept Offset & $c$ & $< \pm 1.0$ ADU \\
Fit Quality & $R^2$ & $> 0.9999$ \\
Linear Threshold Limit & $S_{\mathrm{linear, max}}$ & $\sim 11040$ Net ADU ($12048$ Raw ADU) \\
ADC Clipping Limit & $S_{\mathrm{sat}}$ & $15375$ Net ADU ($16383$ Raw ADU) \\ \hline
\end{tabular}
\end{table}

An unweighted ordinary least squares (OLS) linear regression was performed on the bias-subtracted data within the linear domain ($t_i \le 1.0\text{~s}$):
\begin{equation}
S_{\text{pred}}(t_i) = m \cdot t_i + c
\end{equation}
where $m$ represents the sensor conversion slope (ADU/s) and $c$ represents the residual intercept (ADU). Percentage deviation residuals $\delta_i$ were evaluated across all exposures to quantify non-linearity:
\begin{equation}
\delta_i = \left( \frac{S_{\text{net}}(t_i) - S_{\text{pred}}(t_i)}{S_{\text{pred}}(t_i)} \right) \times 100\%
\end{equation}

As shown in Figure~\ref{fig:sensor_linearity}, signal deviation stays strictly within a $\pm1.5\%$ residual band for $t_i\leq 1$s. The analysis confirms the following sensor performance characteristics:
\begin{itemize}
    \item \textbf{Strict Linearity Domain:} Within $t_i \le 1.0\text{~s}$ ($S_{\text{net}} \le 11{,}040\text{~ADU}$), signal deviation stays strictly within a $\pm 1.5\%$ residual tolerance band, exhibiting an exceptionally high coefficient of determination ($R^2 > 0.9999$).
    \item \textbf{Zero-Point Consistency:} Residual intercept values near $0\text{~ADU}$ after bias subtraction demonstrate that short-exposure micro-shutter timing strictly converges to $t = 0\text{~s}$ without electronic latency artifacts.
    \item \textbf{Upper Saturation:} Onset of non-linearity begins above $11,040$ Net ADU ($12,048$ Raw ADU), with full 14-bit ADC saturation clipping occurring at $15{,}375$ Net ADU ($16,383$ Raw ADU).
\end{itemize}

\begin{authorcontribution}
S.S. made substantial contributions to the conception or design of the work.  S.S., B.A., and P.S. were involved in the acquisition of data. P.S., S.B. and C.G.  were involved in analysis of the data and preparation of figures. S.S. and B.A. were involved in the interpretation of data. S.S., P.S., B.A. and S.B.  wrote the main manuscript text. All authors reviewed the manuscript.
\end{authorcontribution}

\bibliographystyle{spr-mp-sola}
\bibliography{references}

\end{document}